\documentclass[a4paper,12pt]{article}
\usepackage{amsmath,amsfonts,latexsym}
\usepackage{graphicx}
\usepackage{natbib}
\usepackage[hyphens]{url}
\usepackage{hyperref}
\usepackage{chngcntr}
\counterwithout{table}{section}
\newcommand{\bftheta}{\mbox{\boldmath{$\theta$}}}
\newcommand{\bc}{\begin{center}}
\newcommand{\ec}{\end{center}}
\newcommand{\bt}{\begin{tabular}}
\newcommand{\et}{\end{tabular}}
\usepackage{chngcntr}
\counterwithout{table}{section}
\begin{document}
\bc {\large \bf Pseudo-Bayesian Optimal Designs for Fitting Fractional
Polynomial Response Surface Models} \ec
{\large Luzia A.~Trinca}\\
\textit{Unesp, Botucatu, Brazil}

{\large and Steven G.~Gilmour}\\
\textit{King's College London, UK}

\textbf{Summary.} \textsf{Fractional polynomial models are
potentially useful for response surfaces investigations. With the
availability of routines for fitting nonlinear models in
statistical packages they are increasingly being used. However, as
in all experiments the design should be chosen such that the model
parameters are estimated as efficiently as possible. The design
choice for such models involves the known nonlinear models' design
difficulties but \cite{gilmour_trinca_2012b} proposed a methodology
capable of producing exact designs that makes use of the computing
facilities available today. In this paper, we use this methodology
to find Bayesian optimal exact designs for several fractional
polynomial models. The optimum designs are compared to various
standard designs in response surface problems.}


\section{Introduction}

Low order polynomial models are widely used in Response Surface
(RS) methods to approximate the unknown relationship between several
quantitative treatment factors, say $x_1,~x_2,~\ldots,~x_q$ and a continuous
response variable, say $y$ \citep{box_draper_1987}. Despite the
well-known advantages of such approximations, there are occasions
when low order polynomials do not fit the data adequately. Higher
order polynomials may provide a better fit, but they are difficult
to interpret and sometimes can lead to unrealistic predictions.
Gilmour and Trinca (2005) argued that in cases where standard
polynomials are inadequate fractional polynomials (FP) can be very
useful. Fractional polynomial models were inspired by \cite{box_tidwell_1962} transformations in the predictors, that is using
\[
x_i^{(\alpha_i)}=\left\{\begin{array}{cc}
x^{\alpha_i},&\alpha_i\ne0\\
log(x_i),&\alpha_i=0\end{array}\right., 
\]
for $i=1,~2,~\ldots,~q$.
Given the $\alpha_i$s, polynomial models in the transformed
metrics would then be fitted to the response. Since the models
arising are nonlinear in the power parameters ($\alpha_i$) their
practical applications were very much delayed perhaps due to the
computational difficulties involved in estimating these
parameters. 

Royston and Altman (1994) reduced this class of
models allowing rational powers and extending to non functional
marginality. They fitted these models to many data sets from
observational studies showing their flexibility to cope with many
different shapes of relationships. Exploiting the easy availability of
nonlinear least squares routines in statistical packages and the
flexibility of FPs in coping with different asymmetrical shapes of
surfaces, \cite{gilmour_trinca_2005a} showed how they could be
routinely used in RS experiments. They noted, however, in the RS
context the general class of Royston and Altman's models are
difficult to fit because the usual restrictions on experiment
sizes and numbers of factor levels. Low order marginal FP models
are more plausible for RS studies. However even for the marginal FP models
the popular RS designs such as three-level designs and CCDs can be
inefficient to estimate the power parameters. RS designs that take
into account the estimation of the extra power parameters can be
very useful when lack of fit of low order polynomials in the
original factor metrics is suspected. Since such models are
nonlinear, optimal designs for such a task depend on the values of
the unknown parameters. Designs selected considering prior point
parameter values may be very sensitive to the set of values
chosen. A more robust approach to construct optimal designs in
such situations is incorporating prior information on the
parameters into the design by taking the expectation of the design
criterion over the prior distributions. Designs chosen following
this approach are called optimal Bayesian (or pseudo-Bayesian) designs. 

There is a wide
literature on optimum designs for nonlinear models, most of which
is about constructing continuous optimal designs. For Bayesian
designs see, for example, \cite{Atkinson2007}, \cite{pukelsheim_1993} and \cite{chaloner_verdinelli_1995}. Construction of exact
designs for nonlinear models is rarely considered in the
literature. \cite{gilmour_trinca_2012b} discussed the difficulties
related to obtaining optimum exact designs for experiments
exploring non-linear relationships between the response and its
predictors and presented an approach to implement exact Bayesian
designs. 

The aim of this paper is to present the designs obtained
by using this approach for a wide range of practical situations
that may arise in the RS context. We restrict attention
to the cases of one and two factors, due to computational issues, and believe these findings can contribute
to the construction of designs when more factors are used. In the
next two sections the FP models are explicitly presented and
reparameterizations to facilitate interpretation and design
construction are proposed. Expressions for the variances of the
non-linear least squares parameter estimators are also developed.
In section \ref{criterion} the design criterion we use is
presented; in section \ref{priors} practical issues concerning the
choice of parameter prior distributions are presented; in section
\ref{methods} a brief summary of the methodology to obtain exact
designs is presented and in section \ref{examples} several exact
optimum designs are presented and compared to standard designs. The paper concludes with some practical advice in section \ref{advice}.

\section{Fractional polynomial models}

In RS experiments usually resources are limited and so it is
convenient to use not too many different levels of the factors.
Consequently useful FP models are the first and second order
models which are polynomial models under the
transformed metric ($x^{(\alpha)}$). For one factor, first and
second order FP are
\begin{equation}
y = \beta_0 + \beta_{1}x^{(\alpha)} + \epsilon \label{fofp}
\end{equation}
and
\begin{equation}
y = \beta_0 +\beta_{1}x^{(\alpha)} + \beta_{11}\{x^{(\alpha)}\}^2
+ \epsilon, \label{sofp}
\end{equation}
respectively, where $x^{(\alpha)}$ is \begin{eqnarray}
 x^{(\alpha)} = \left\{\begin{array}{ll}
x^{\alpha}, &
\alpha\neq 0;\\
                                            \log{x},      &
\alpha=0\end{array}\right. \label{xalpha}
\end{eqnarray}
and $\epsilon \sim N\left( 0,\sigma^2\right)$. Note that model
(\ref{fofp}) is a reparameterization of Mitscherlich growth model. Note also that the models are restrict to $x>0$ so
the $x$ levels should not be coded by using the usual RS coding.
We find it useful to code $x$ such that $x \in [x_{min}; 1]$. Let the
models in (\ref{fofp}) and (\ref{sofp}) be written in the general
nonlinear form $y=\eta( x,\bftheta)+\epsilon$ where $\bftheta$ is
a ($p\times 1$) vector of unknown parameters. It can be shown
that, for $n$ independent observations ${\bf y}$, an approximation
for the asymptotic covariance matrix of $\hat\bftheta$, the maximum likelihood or nonlinear least squares estimator
of $\bftheta$, given $\bftheta$, is
\begin{equation}
{\bf M}({\bf X},\bftheta)^{-1}=\left[{\bf F}({\bf
X},\bftheta)^\prime{\bf F}({\bf X},\bftheta)\right]^{-1}\sigma^2,
\label{covnl}
\end{equation} where
${\bf F}({\bf X},\mbox{\boldmath $\theta$})$ is the $n\times p$
matrix expansion of
\begin{eqnarray}{\bf f}(
x,\bftheta)=\left\{\frac{\partial\eta( x,\bftheta)}
{\partial\theta_1}~\frac{\partial\eta( x,\bftheta)}
{\partial\theta_2}~...~\frac{\partial\eta( x,\bftheta)}
{\partial\theta_p}\right\} \mid{_{\bftheta=\bftheta_o}}.
\label{partial}
\end{eqnarray}
The problem of selecting an exact
design is that of selecting the number of different levels of $x$,
say $k$, the levels themselves, say $x_i$, and their replication
$n_{i}$ $(i=1, 2,\ldots,k)$, such that some property of the
covariance matrix (or its inverse), the design criterion,
$\phi({\bf M}({\bf X},\bftheta))$, is optimized. The covariance
matrix is a function of the true unknown $\bftheta$ values and
consequently the optimum design usually depends on these. For a
particular parameter value, say $\mbox{\boldmath $\theta$}_o$, the
optimization of $\phi({\bf M}({\bf X},\mbox{\boldmath $\theta$}))$
yields a locally optimum design which may not be robust for other
values of $\bftheta$. A Bayesian approach to make the designs more
robust in a wide range of the parametric space is very useful.
Here we apply the methodology proposed in \cite{gilmour_trinca_2012b} to obtain optimum exact designs for the models in
(\ref{fofp}) and (\ref{sofp}). This approach consists in
optimizing
\begin{equation}
\varphi({\bf X})=E_{\bftheta}\{\phi({\bf M}({\bf X},\bftheta))\}=
\int_{\bftheta\in\Theta} \phi({\bf M}({\bf X},\bftheta))
p(\bftheta) d\bftheta,\label{crit}
\end{equation}
where $p(\bftheta)$ is a prior probability distribution for
$\bftheta$. Independent prior distributions among the parameters
are desired and we propose some reparameterisation of the models in
order to ensure direct interpretation of the parameters to be
estimated. Such reparameterisation will be described in Section
\ref{modelrep}. However, even for independent prior
distributions an algebraic solution for the criterion in (\ref{crit})
is usually not available. The proposed approach suggests
approximating the criterion by sampling $r$ points from each
parameters's prior distribution. Thus, the approximated criterion
becomes
\begin{equation}\tilde{\varphi}(X)=\frac{1}{r}\sum_{j=1}^r\phi({\bf
M}({\bf X},\tilde\bftheta_j)),\label{critapp}\end{equation} where
$\tilde\bftheta_1$, $\tilde\bftheta_2$, $\ldots$,
$\tilde\bftheta_r$ are the sample parameters simulated. For each
design {\bf X} the quantity in (\ref{critapp}) is evaluated and
the {\bf X} that optimize it is the optimum design.

\section{Model reparameterizations} \label{modelrep} For
parameters that represent quantities of practical interest which
are directly interpretable the task of specifying their prior
distributions is not too difficult. Further, if such
quantities are independent from each other, the process of specifying priors is very much
simplified. However for the FP models the interpretation of the
regression parameters $\mbox{\boldmath{$\beta$}}$ depends on the
value of $\alpha$. Reparametrisations of the models in terms of
quantities which are more directly interpretable are required.

A quantity of particular interest in RS experiments is the change
in the mean response as $x$ changes from its smallest, say
$x_{min}$, to the largest, say $x_{max}$, levels. Lets call this
quantity $\gamma_1$. For the first order FP model (\ref{fofp}),
and coding $x$ such that $x \in [x_{min};1]$, $\gamma_1$ is
\begin{eqnarray*}
\gamma_1 =  E\left[y(1)-y(x_{min})\right] =
\beta_1\left(1^{(\alpha)}-x_{min}^{(\alpha)}\right)\Rightarrow
\beta_1=\gamma_1\frac{1}{1^{(\alpha)}-x_{min}^{(\alpha)}}.
\end{eqnarray*}
Hence the reparameterized model in terms of $\gamma_1$ is
\begin{equation}
y = \beta_0 +
\gamma_1\frac{x^{(\alpha)}}{1^{(\alpha)}-x_{min}^{(\alpha)}} +
\epsilon \label{fofpr}.
\end{equation}

For a curved relationship in the transformed metric
$x^{(\alpha)}$, besides $\gamma_1$, the difference between the
mean response for $x$ at its extreme values and at the center, say
$\gamma_{11}$, is also of interest. For the second order FP model
(\ref{sofp}), $\gamma_{1}$ and $\gamma_{11}$ are, respectively,
\begin{eqnarray*}
\gamma_{1}=\beta_1\left(1^{(\alpha)}-x_{min}^{(\alpha)}\right)+\beta_{11}\left(1^{2(\alpha)}-x_{min}^{2(\alpha)}\right)
\end{eqnarray*}
and
\begin{eqnarray*}
\gamma_{11} =
E\left[\frac{y(x_{min})+y(1)}{2}-y\left(x_c^{\star}\right)\right]=\beta_{11}\frac{\left(1^{(\alpha)}-x_{min}^{(\alpha)}\right)^2}{4},\\
\end{eqnarray*}
where $x_c^{\star}=\frac{1^{(\alpha)}+x_{min}^{(\alpha)}}{2}$.
These yield
\begin{eqnarray*}\beta_{11}=4(1^{(\alpha)}-x_{min}^{(\alpha)})^{-2}\gamma_{11}\end{eqnarray*}
and
\begin{eqnarray*}\beta_1=(1^{(\alpha)}-x_{min}^{(\alpha)})^{-1}\gamma_1
-4(1^{(\alpha)}+x_{min}^{(\alpha)})(1^{(\alpha)}-x_{min}^{(\alpha)})^{-2}\gamma_{11}.\end{eqnarray*}
Hence the reparameterized second order FP is
\begin{equation}
y=\beta_0+\gamma_1\frac{x^{(\alpha)}}{1^{(\alpha)}-x_{min}^{(\alpha)}}+
4\gamma_{11}\frac{1}{(1^{(\alpha)}-x_{min}^{(\alpha)})^2}\left[x^{2(\alpha)}-(1^{(\alpha)}
+x_{min}^{(\alpha)})x^{(\alpha)} \right]+\epsilon \label{sofpr}
\end{equation}
Thus, in practical terms, reasonable prior distributions can be
specified independently for $\gamma_1$ and $\alpha$ in model
(\ref{fofpr}) and for $\gamma_1$, $\gamma_{11}$ and $\alpha$ in
model (\ref{sofpr}).

{\it Variances for the parameter estimators for the first order FP
model}

For model (\ref{fofpr}) the approximation in (\ref{partial})
yields
\begin{eqnarray*}
\frac{\partial
E(y)}{\partial{\gamma_1}}=\frac{x^{(\alpha)}}{1^{(\alpha)}-x_{min}^{(\alpha)}}=
\left\{\begin{array}{ll}\frac{1}{1-x_{min}^{\alpha}} x^{\alpha}, &  \alpha\neq0\\
-\frac{1}{\ln x_{min}} \ln x,
&\alpha=0\end{array}\right.\mbox{\normalsize ~~~~~~~~~~~~and}
\end{eqnarray*}
\begin{eqnarray*}
\frac{\partial
E(y)}{\partial{\alpha}}&=&\gamma_1\frac{x^{(\alpha)}\left((1^{(\alpha)}-x_{min}^{(\alpha)})\ln
x +x_{min}^{(\alpha)}\ln
x_{min}\right)}{(1^{(\alpha)}-x_{min}^{(\alpha)})^2}\\&=&
\left\{\begin{array}{ll}\frac{\gamma_1}{(1-x_{min}^{\alpha})^2}\left[
(1-x_{min}^{\alpha})x^{\alpha}\ln x +x_{min}^{\alpha}\ln
x_{min}x^{\alpha}\right]  , &  \alpha\neq0\\
\frac{\gamma_1}{\ln^2 x_{min}} \left[ \ln x_{min}\ln x(\ln
x_{min}-\ln x)\right], &\alpha=0\end{array}\right..
\end{eqnarray*}
From these the matrices ${\bf F}({\bf
X},\mbox{\boldmath{$\theta$}})$ and ${\bf M}({\bf
X},\mbox{\boldmath{$\theta$}})$, both of dimension $3\times 3$,
can be constructed and the diagonal elements of ${\bf M}({\bf
X},\mbox{\boldmath{$\theta$}})^{-1}$ obtained by using Maple, for
example. Writing ${\bf M}({\bf X},\mbox{\boldmath{$\theta$}})$ as
${\bf M}$ (for short) results in
\[
Var(\hat\gamma_1\mid\mbox{\boldmath
$\theta$})=\frac{1}{d}\left(n[M]_{33}-[M]_{13}^2\right)
\]
and
\[
Var(\hat\alpha\mid\mbox{\boldmath
$\theta$})=\frac{1}{d}\left(n[M]_{22}-[M]_{12}^2\right),
\]
where
$
d=n[M]_{22}[M]_{33}-n[M]_{23}^2-[M]_{12}^2[M]_{33}+2[M]_{12}[M]_{13}[M]_{23}-[M]_{13}^2[M]_{22}\\
$ and $[M]_{ij}$ is the $ij^{th}$ element of the matrix ${\bf
M}({\bf X},\mbox{\boldmath{$\theta$}})$. It is possible to show
that $d$, $[M]_{33}$ and $[M]_{13}^2$ are proportional to
$\gamma_1^2$ and so $Var(\hat\gamma_1\mid\mbox{\boldmath
$\theta$})$ does not depend on $\gamma_1$.

{\it Variances for the parameter estimators for the second order FP model}

For model (\ref{sofpr}) the approximation in (\ref{partial})
yields
\begin{eqnarray*}
\frac{\partial
E(y)}{\partial{\gamma_1}}=\frac{x^{(\alpha)}}{1^{(\alpha)}-x_{min}^{(\alpha)}}=
\left\{\begin{array}{ll}\frac{1}{1-x_{min}^{\alpha}} x^{\alpha}, &  \alpha\neq0\\
-\frac{1}{\ln x_{min}} \ln x, &\alpha=0\end{array}\right.,
\end{eqnarray*}

\begin{eqnarray*}
\frac{\partial
E(y)}{\partial{\gamma_{11}}}&=&4\frac{\left\{x^{(\alpha)}\right\}^2-(1^{(\alpha)}+x_{min}^{(\alpha)})x^{(\alpha)}}{(1^{(\alpha)}-x_{min}^{(\alpha)})^2}
\\&=&\left\{\begin{array}{ll}\frac{4}{(1-x_{min}^{\alpha})^2}\left(
x^{2\alpha}-(1+x_{min}^{\alpha})x^{\alpha}\right)
, &  \alpha\neq0\\
\frac{4} {\ln^2x_{min}}\left(\ln^2 x-\ln x_{min}\ln x\right) , &
\alpha=0\end{array}\right. \mbox{\normalsize ~~~~~~~~~~~~and}
\end{eqnarray*}

\begin{eqnarray*}
\frac{\partial E(y)}{\partial{\alpha}}&=&\frac{1}{
(1^{(\alpha)}-x_{min}^{(\alpha)})^3}\times\left[1^{(\alpha)}\left(\left(
\gamma_1(1-2x_{min}^{(\alpha)})
-4\gamma_{11}\right)x^{(\alpha)}\ln
x\right.\right.+\\
& &\left.\left(\gamma_1-12\gamma_{11}\right)x_{min}^{(\alpha)}\ln
x_{min} x^{(\alpha)}+ 8\gamma_{11}\left\{x^{(\alpha)}\right\}^2\ln
x\right)-\\
& &\gamma_1\left\{x_{min}^{(\alpha)})\right\}^2\ln x_{min}x^{(\alpha)}-\gamma_1\left\{x_{min}^{(\alpha)}\right\}^2x^{(\alpha)}\ln x+\\
& & 4\gamma_{11}\left\{x_{min}^{(\alpha)}\right\}^2x^{(\alpha)}\ln
x+4\gamma_{11}(2-x_{min}^{(\alpha)})x_{min}^{(\alpha)}\ln x_{min}
x^{(\alpha)}\\
& &\left.- 8\gamma_{11}
x_{min}^{(\alpha)}\left\{x^{(\alpha)}\right\}^2\ln x\right] \\
&=&\left\{\begin{array}{ll}\frac{1}{(1-x_{min}^{\alpha})^3}\times\left[k_1x^{\alpha}\ln
x + k_2x^{\alpha} + k_3x^{2\alpha}
 +k_4x^{2\alpha}\ln x \right]&\alpha\neq0\\
\frac{1}{\ln^3 x_{min}}\times\left[k_5\ln x+k_6\ln^2
x+k_7\ln^3x\right]&\alpha=0,\end{array}\right.
\end{eqnarray*}
where \begin{eqnarray*}
k_1&=&\gamma_1(1-x_{min}^{\alpha})^2-4\gamma_{11}(1-x_{min}^{2\alpha});\\
 k_2&=&\left(
\gamma_1(1-x_{min}^{\alpha})-4\gamma_{11}(3+x_{min}^{\alpha})\right)x_{min}^{\alpha}\ln
x_{min}; \\k_3&=&8\gamma_{11}x_{min}^{\alpha}\ln x_{min}\\
k_4&=&8\gamma_{11}(1-x_{min}^{\alpha});\\
k_5&=&(4\gamma_{11}+\gamma_1)\ln^3x_{min};\\
k_6&=&\left(4\gamma_{11}(1-3\ln^2 x_{min})-\gamma_1(1+\ln^2
x_{min})\right) \mbox{\normalsize and} \\
k_7&=&8\gamma_{11}\ln x_{min}. \end{eqnarray*} As for the first
order model, the matrices ${\bf F}({\bf
X},\mbox{\boldmath{$\theta$}})$ and ${\bf M}({\bf
X},\mbox{\boldmath{$\theta$}})$, both of dimension $4 \times 4$,
can be constructed and the diagonal elements of ${\bf M}({\bf
X},\mbox{\boldmath{$\theta$}})^{-1}$ obtained by using a
mathematical symbolic package. The variances of interest are
\begin{eqnarray}
Var(\hat\gamma_1\mid\mbox{\boldmath
$\theta$})&=&\frac{1}{d}\left(n[{M}]_{33}[{M}]_{44}-n[{M}]_{34}^2-[{
M}]_{13}^2[{M}]_{44}+ 2[{
M}]_{13}[{M}]_{14}[{M}]_{34}-\right.\nonumber\\&&\left.[{
M}]_{14}^2[{ M}]_{33}\right)\sigma^2, \label{vg1}
\end{eqnarray}
\begin{eqnarray}
Var(\hat\gamma_{11}\mid\mbox{\boldmath
$\theta$})&=&\frac{1}{d}\left(n[{ M}]_{22}[{ M}]_{44}-n[{
M}]_{24}^2-[{ M}]_{12}^2[{ M}]_{44} + 2[{ M}]_{12}[{ M}]_{14}[{
M}]_{24}-\right.\nonumber\\& &\left.[{ M}]_{14}^2[{
M}]_{22}\right) \sigma^2\label{vg11}
\end{eqnarray}
and
\begin{eqnarray}
Var(\hat\alpha\mid\mbox{\boldmath $\theta$})&=&\frac{1}{d}\left(
n[{ M}]_{22}[{ M}]_{33}-n[{ M}]_{23}^2-[{ M}]_{12}^2[{ M}]_{33}+[{
M}]_{12}[{ M}]_{13}[{ M}]_{23} -\right.\nonumber\\
& &\left.2[{ M}]_{13}^2[{ M}]_{22}\right)\sigma^2, \label{valpha}
\end{eqnarray}
where
\begin{eqnarray*}
d&=&n[{ M}]_{22}[{ M}]_{33}[{ M}]_{44}-n[{ M}]_{22}[{ M}]_{34}^2-n[{ M}]_{23}^2[{ M}]_{44}-n[{ M}]_{24}^2[{ M}]_{33}+\\
& & 2n[{ M}]_{23}[{ M}]_{24}[{ M}]_{34}-[{ M}]_{12}^2[{ M}]_{33}[{
M}]_{44}+2[{ M}]_{12}[{ M}]_{23}[{ M}]_{13}[{ M}]_{44}+\\& &[{
M}]_{12}^2[{ M}]_{34}^2-2[{ M}]_{12}[{ M}]_{23}[{ M}]_{14}[{ M}]_{34}-2[{ M}]_{12}[{M}]_{24}[{ M}]_{13}[{ M}]_{34}+\\
& &2[{ M}]_{12}[{ M}]_{24}[{M}]_{14}[{ M}]_{33}-[{ M}]_{22}[{
M}]_{13}^2[{ M}]_{44}+[{ M}]_{13}^2[{
M}]_{24}^2+\\
& &2[{ M}]_{13}[{ M}]_{22}[{ M}]_{14}[{ M}]_{34}-2[{ M}]_{13}[{
M}]_{24}[{ M}]_{14}[{ M}]_{23}-\\& &[{ M}]_{22}[{
M}]_{14}^2[{ M}]_{33}+[{ M}]_{14}^2[{ M}]_{23}^2\\
&=&g({\bf X},\gamma_1,\gamma_{11},\alpha),
\end{eqnarray*}
and $[M]_{ij}$ is the $ij^{th}$ element of the matrix ${\bf
M}({\bf X},\mbox{\boldmath{$\theta$}})$. For this model further
simplifications are difficult but we can see all the variances
depend on the true values of all parameters.
\section{The design criterion}
\label{criterion} There are several suitable design criteria
available in the optimal design theory, the most popular being $D$
and $A$ based criteria. In the RS context usually the experimenter
is more interested in answering specific questions about some
quantities. For this reason we think $A$ based criteria are more
appropriate. Here we use a version of weighted
$A_s$-optimality. Weighted $A_s$-optimality represents the
weighted average variances of a subset of parameters considered of
interest for estimation. The reparameterised form of models
presented in (\ref{fofpr}) and (\ref{sofpr}) are well in line with
this thinking, interest being mainly in estimating the
regression coefficients ($\gamma_1$ and $\gamma_{11}$), but also
allowing good estimation of $\alpha$. Using the weighted $A_s$
criterion the design construction is tailored such that all the
parameters are estimated precisely with the flexibility of some
parameters being given relatively higher priorities. This
flexibility compensates in some way the lack of scale invariance
of $A$ optimum designs and is thus well suited for FP models.
Hence the design criterion we use is given by $\varphi(\mathbf{X})
=
E_{\bftheta}[\mbox{tr}\{\mathbf{WV}(\hat{\bftheta}|\bftheta,\mathbf{X})\}]$
where {\bf W} is a diagonal matrix with the weights for each
parameter. The criterion function $\varphi(\mathbf{X})$  will be approximated by
\begin{equation}\tilde\varphi(\mathbf{X}) =
\sum_{j=1}^r[\mbox{tr}\{\mathbf{WM}(\mathbf{X},\tilde\bftheta_j)\}].\label{trace}\end{equation}
For our models the first element of {\bf W} is set to zero since
we are not interested in the models' intercept. Since for model
(\ref{fofp}) we usually want very precise estimate of $\gamma_1$
and good estimate of $\alpha$ but for $\gamma_1\rightarrow 0$ no
matter what $\alpha$ estimate is it seems reasonable to make the
elements of {\bf W} dependent on $\gamma_1$. A similar argument can
be applied to the model in (\ref{sofp}) and thus, in this case,
we make the elements of {\bf W} dependent on the regression
parameters prior values. Details of how this is implemented is
presented in the next section.

\section{Choosing priors and weights for the parameters}
\label{priors} For fitting FP models \cite{gilmour_trinca_2005a}
argued that the $\alpha$ estimate could satisfactorily be rounded to some
number in the set
$\left\{-3,-2,-1,-\frac{1}{2},-\frac{1}{3},0,\frac{1}{3},\frac{1}{2},1,2,3\right\}$.
Now we argue that a probability mass on the values of this set or an even more restricted set can provide a reasonable prior
distribution for $\alpha$. Here we use the set
$\left\{-2,-1,-\frac{1}{2},0,\frac{1}{2},1,2\right\}$. High
probabilities can then be assigned to negative values if it is
believed the response curve should be asymptotic as
$x\rightarrow\infty$, high probabilities can be assigned to positive values if the
experimenter suspects it continues to increase or decrease as
$x\rightarrow\infty$, uniform probabilities give maximum
uncertainty and probabilities can be concentrated near 0 if a log
transformation is suspected to be adequate, etc.

For the regression parameters ($\gamma_1$ and $\gamma_{11}$) the
natural prior distribution is the Normal density. For the first
order FP model prior mean values for $\gamma_1$ should be positive
if the response is believed to increase with $x$ in the
transformed metric and negative if it is believed to decrease. For the second order FP
model prior mean values for $\gamma_{11}$ should be large if it is
suspected there is a turning point within the experimentation
region, positive or negative, depending on the turning point being
a maximum or a minimum. The prior variances for $\gamma_1$ and
$\gamma_{11}$ may be small or large depending on the experimenter's
uncertainty about these quantities.

For the models considered in this paper common sense reasoning
allows us to stipulate maximum target variances for the estimates
of each parameter, that is
\[Var(\hat\gamma_1)\le\frac{1}{4}\gamma_1^2\]
\[Var(\hat\gamma_{11})\le\frac{1}{4}\gamma_{11}^2\]
and
\[Var(\hat\alpha)\le\frac{1}{4} \mbox{\normalsize ~~~(considering the prior values for $\alpha$ proposed above)}.\]
Thus, for the first order FP model, the target variances imply the
weights $w_1$ and $w_2\gamma_1^2$ for $\gamma_1$ and
$\alpha$, respectively, with $w_1=w_2=1$ in case of equal targets.
Hence, the diagonal elements of the matrix {\bf W} in
(\ref{trace}) are \begin{eqnarray} W_{11}=0, \mbox{\hspace{0.5cm}}
W_{22}=\frac{w_1}{w_1+w_2\gamma_1^2} \mbox{\normalsize ~~and~~}
W_{33}=\frac{w_2\gamma_1^2}{w_1+w_2\gamma_1^2}\label{wfofp}.\end{eqnarray}
As we have more interest in estimating $\gamma_1$ we suggest using
$w_2=1$ and $w_1>1$. In doing so, for $\gamma_1$ close to zero the
weight for $\alpha$ is very small which is quite reasonable since
for $\gamma_1=0$ we would not be interested in $\alpha$.
Similarly, for the second order FP model, the target variances
imply weights $w_1\gamma_{11}^2$, $w_2\gamma_1^2$ and
$w_3\gamma_1^2\gamma_{11}^2$ for $\gamma_1$, $\gamma_{11}$ and
$\alpha$, respectively, with $w_1=w_2=w_3=1$ in case of equal
targets. Hence, the diagonal elements of the matrix {\bf W} now
are \begin{eqnarray}W_{11}&=&0,\mbox{\hspace{0.5cm}}
W_{22}=\frac{w_1\gamma_{11}^2}{w_1\gamma_{11}^2+w_2\gamma_1^2+w_3\gamma_1^2\gamma_{11}^2},\mbox{\hspace{1cm}}\nonumber\\
W_{33}&=&\frac{w_2\gamma_{1}^2}{w_1\gamma_{11}^2+w_2\gamma_1^2+w_3\gamma_1^2\gamma_{11}^2}
\mbox{\normalsize ~~~~~~~~and}\nonumber\\
W_{44}&=&\frac{w_3\gamma_1^2\gamma_{11}^2}{w_1\gamma_{11}^2+w_2\gamma_1^2+w_3\gamma_{1}^2\gamma_{11}^2}.\label{wsofp}\end{eqnarray}
Here we suggest using $w_1>w_2>w_3=1$.
\section{Methods for obtaining optimum designs}
\label{methods}
 
\cite{gilmour_trinca_2012b} showed how practical
optimum Bayesian exact designs for nonlinear models could be
obtained using the existing computer facilities. The difficulties
of differentiating $\eta({\bf x})$ and inverting variances
matrices numerically can be overcome by using a computer algebra
package and the expected design criterion can be approximated
numerically by sampling from the prior distributions. These two
well known techniques can be implemented in the search for optimum
designs in two ways: by using an algorithm of exchange type or by
performing a complete search over two combined grids, one for
design points (levels) and the other for design weights (level
replications). For very fine grids the resulting design will be
very close to the continuous design.  The complete search can also
be used for exact designs if the degree of fineness of the design
weight grid is set to a probability which relative to the size of
the experiment (total number of runs) represents one run. For
example, for an experiment with 20 runs the degree of fineness of the
design weight grid smaller than 0.05 is of no use. For small
designs and models with few parameters it is quite reasonable to
do the complete search. For large designs or large models the
exchange algorithm is more reasonable and perhaps the only viable
alternative. It consists of improving an initial design by
exchanging design points with points from a candidate set. To
increase the chance of getting the optimum design the exchange
method is repeated for a number of different initial design
(tries). The exchange algorithm has been frequently used for
searching optimum designs for linear models. For nonlinear models
a detailed description is given by \cite{gilmour_trinca_2012b}.

We use an example to compare the performance of the exchange and
the complete search methods for a small problem, a 12-run design
in the $x$ range [0.1,~1] for the first order FP model given in
(\ref{fofpr}). The prior parameter distributions were
$\gamma_1\sim N(2.5;1.5^2)$ and a seven-point prior for $\alpha$,
with probabilities 0.15, 0.25, 0.25, 0.15, 0.10, 0.07 and 0.03 for
$\alpha =-2$, $-1$, $-0.5$, 0, 0.5, 1 and 2, respectively, and
used equal target variances for both parameters ($w_1=w_2=1$ in
(\ref{wfofp})). The number of points sampled from $\gamma_1$'s
prior was $r=200$ and $\alpha$ values in the set above were
replicated accordingly. For the complete search we used a fineness
of 1/12 for the replication grid in order to obtain the exact
design for an experiment based on 12 runs. Initially we used a
fineness grid of 0.01 for the $x$ levels for both methods. For the
complete search, because computer time was quite an issue, only 3
or 4 levels designs were allowed for (as there are 3 parameter to
be estimated we expected the local optimum design would use 3
levels at least). The design obtained by complete search was 3
replicates at 0.1, 4 at 0.19, 2 at 0.52 and 3 at 1.0 with
$\tilde\varphi=0.4444665$. A finer grid of 0.001 around the levels
0.19 and 0.52 (now fixing one level at 0.1 and one at 1.0)
produced a slightly superior design ($\tilde\varphi=0.444479$)
being 3 replicates at 0.1, 4 at 0.189, 2 at 0.521 and 3 at 1. The
exchange version (3 tries) produced very similar design: 3
replicates at 0.1, 4 at 0.19, 1 at 0.52, 1 at 0.53 and 3 at 1.0
with $\tilde\varphi=0.4444691$. Rounding the levels with 1
replication to 0.525 resulted in some improvement
($\tilde\varphi=0.4444753$). Although this resulted in a quite
satisfactory design we run the exchange version again using the
same finer grid around the middle levels as before. The resulting
design was 3 replicates at 0.1, 2 at 0.189, 2 at 0.190, 1 at
0.522, 1 at 0.523 and 3 at 1.0 with $\tilde\varphi=0.4444817$.
Again, rounding the levels, that is, 4 replications at 0.1895 and
2 at 0.5225 produced the best design with
$\tilde\varphi=0.4444843$. We could continue making the grid finer and
possibly obtain the optimum exact design. However the gain at each
step is very small and we believe that for practical designs a
simple exchange search produces quite efficient designs. Note that
because the restriction on the fineness of the grid and the number
of levels allowed for the complete search, for the same fineness
the exchange version produce superior designs. We thought further
comparisons with using complete search allowing 5 levels or more
were not fruitful since the computer time spent in the cases
studied were about 6 times bigger than the exchange version. These
results indicate that the exchange algorithm can successfully be
used in practice and from now on we restrict to results based on
this method only.

\section{Illustrations}\label{examples}
In this section we present several optimum exact designs for
first and second FP models, varying the prior distributions, the
weight patterns and the size of the designs. Firstly, we present
some results to give an idea of the sensitivity of the designs to
the parameter weights entered in the criterion as well as the
sensitivity of the designs with respect to the prior parameter
distributions used for the regression parameters. Secondly we
compare the optimum designs to some popular designs used in
practice and show how the designs vary subject to $\alpha$ prior
knowledge. \subsection{Design sensitivity to the regression
parameter priors and weight patterns}

\noindent \textit{First Order Model}

\begin{table}
\caption{Effect of changing the $\gamma_1$ prior distributions and
the parameter weight pattern on the exact optimum designs for
fitting a first order FP model.
$\alpha=(-2,~-1,~-1/2,~0,~1/2,~1,~2)$ with $
P(\alpha)=(0.15,~0.25,~0.25,~0.15,~0.10,~0.07,~0.03)$.}
\label{tabw1}
\begin{center}
\begin{small}
\begin{tabular}{cc}
 \hline
Weight pattern& $\begin{array}{ccccc}&&Design&&\end{array}$\\
\hline &$\begin{array}{ccccc}&&{\bf n=12}&&\end{array}$\\
&$\begin{array}{ccccc}&&{\bf \gamma_1\sim N(2.5;1.5^2)}&&\end{array}$\\
$w_1=w_2=1$&$\left\{\begin{array}{ccccc}x&0.1000&0.1895&0.5225&1.0000\\
n_i&3&4&2&3\end{array}\right.$\\
$w_1=10$; $w_2=1$&$\left\{\begin{array}{ccccc}x&0.1000&0.1875&0.5700&1.0000\\
n_i&3&4&2&3\end{array}\right.$\\
&$\begin{array}{ccccc}&&{\bf \gamma_1\sim N(2.5;0.5^2)}&&\end{array}$\\
$w_1=w_2=1$&$\left\{\begin{array}{ccccc}x&0.1000&0.1900&0.5300&1.0000\\
n_i&3&4&2&3\end{array}\right.$\\
$w_1=10$; $w_2=1$&$\left\{\begin{array}{ccccc}x&0.1000&0.1875&0.5750&1.0000\\
n_i&3&4&2&3\end{array}\right.$\\
&$\begin{array}{ccccc}&&{\bf \gamma_1\sim N(-2.5;1.5^2)}&&\end{array}$\\
$w_1=w_2=1$&$\left\{\begin{array}{ccccc}x&0.1000&0.1900&0.5400&1.0000\\
n_i&3&4&2&3\end{array}\right.$\\
$w_1=10$; $w_2=1$&$\left\{\begin{array}{ccccc}x&0.1000&0.1875&0.5750&1.0000\\
n_i&3&4&2&3\end{array}\right.$\\

 \hline
&$\begin{array}{ccccc}&&{\bf n=20}&&\end{array}$\\
&$\begin{array}{ccccc}&&{\bf \gamma_1\sim N(2.5;1.5^2)}&&\end{array}$\\
$w_1=w_2=1$&$\left\{\begin{array}{ccccc}x&0.1000&0.1910&0.5600&1.0000\\
n_i&5&8&3&4\end{array}\right.$\\
$w_1=10$; $w_2=1$&$\left\{\begin{array}{ccccc}x&0.1000&0.1870&0.5600&1.0000\\
n_i&6&6&3&5\end{array}\right.$\\
&$\begin{array}{ccccc}&&{\bf \gamma_1\sim N(2.5;0.5^2)}&&\end{array}$\\
$w_1=w_2=1$&$\left\{\begin{array}{ccccc}x&0.1000&0.1900&0.5250&1.0000\\
n_i&4&7&4&5\end{array}\right.$\\
$w_1=10$; $w_2=1$&$\left\{\begin{array}{ccccc}x&0.1000&0.1880&0.5700&1.0000\\
n_i&6&6&3&5\end{array}\right.$\\
&$\begin{array}{ccccc}&&{\bf \gamma_1\sim N(-2.5;1.5^2)}&&\end{array}$\\
$w_1=w_2=1$&$\left\{\begin{array}{ccccc}x&0.1000&0.1900&0.5300&1.0000\\
n_i&4&7&4&5\end{array}\right.$\\
$w_1=10$; $w_2=1$&$\left\{\begin{array}{ccccc}x&0.1000&0.1900&0.5670&1.0000\\
n_i&6&6&3&5\end{array}\right.$\\
\hline
\end{tabular}
\end{small}
\end{center}
\end{table}

In Table \ref{tabw1} we show the designs for a few prior
distributions for $\gamma_1$ and weight patterns. The prior for
$\alpha$ did not vary and is the same used in the previous section.
 We consider experiments using
12 and 20 runs. As expected the differences among the designs from
using different priors for $\gamma_1$ are very small. Changing the
weight patterns causes bigger differences: giving higher priority
to $\gamma_1$, in all cases, we note an attempt to move the middle
levels to the extremes as expected. For $n=12$, as the replication
on each level is already critical the replication pattern does not
change but the middle levels become closer to the extremes. For
$n=20$ the replication pattern does change with some from the
middle levels migrating to the extremes.

\noindent \textit{Second Order Model}

\begin{table}
\caption{Effect of changing the $\gamma_1$ and $\gamma_{11}$ prior
distributions and the parameter weight pattern on the exact
optimum designs (n=20) for fitting a second order FP model. $
\alpha=(-2,~-1,~-1/2,~0,~1/2,~1,~2)$ with $
P(\alpha)=(0.15,~0.25,~0.25,~0.15,~0.10,~0.07,~0.03)$.}
 \label{tabw2}
\begin{center}
\begin{small}
\begin{tabular}{cc}
 \hline
Weight pattern& $\begin{array}{ccccc}&&Design&&\end{array}$\\
\hline 
&$\begin{array}{ccccc}&&{\bf \gamma_1\sim N(1;0.5^2)}~{\mbox{\normalsize and}}~{\bf \gamma_{11}\sim
N(-2.5;1.5^2)}&&\end{array}$\\\\
$w_1=w_2=w_3=1$&$\left\{\begin{array}{ccccccc}x&0.1&0.1400&0.1742&0.5100&0.6200&1\\
n_i&3&1&5&6&1&4\end{array}\right.$\\\\
$\begin{array}{c}w_1=w_2=10;\\w_3=1\end{array}$&$\left\{\begin{array}{cccccc}x&0.1&0.1685&0.5122&0.5800&1\\
n_i&3&6&6&1&3\end{array}\right.$\\\\
$\begin{array}{c}w_1=100;\\ w_2=10;~w_3=1\end{array}$&$\left\{\begin{array}{ccccc}x&0.1&0.1700&0.5237&1\\
n_i&3&6&7&4\end{array}\right.$\\\\

&$\begin{array}{ccccc}&&{\bf \gamma_1\sim N(1;0.2^2)}~{\mbox{\normalsize and}}~{\bf \gamma_{11}\sim
N(-2.5;0.5^2)}&&\end{array}$\\\\
$w_1=w_2=w_3=1$&$\left\{\begin{array}{cccccccc}x&0.1&0.1300&0.2310&0.4825&0.6110&0.7760&1\\
n_i&5&3&3&2&1&1&5\end{array}\right.$\\\\
$\begin{array}{c}w_1=w_2=10;\\w_3=1\end{array}$&$\left\{\begin{array}{cccccccc}x&0.1&0.1379&0.2064&0.5329&0.5801&0.7561&1\\
n_i&6&2&2&2&1&1&6\end{array}\right.$\\\\
$\begin{array}{c}w_1=100;\\ w_2=10;~w_3=1\end{array}$&$\left\{\begin{array}{cccccccc}x&0.1&0.1386&0.1798&0.1811&0.5654&0.6699&1\\
n_i&7&1&1&1&1&3&6\end{array}\right.$\\\\

&$\begin{array}{ccccc}&&{\bf \gamma_1\sim N(3;1.5^2)}~{\mbox{\normalsize and}}~{\bf \gamma_{11}\sim
N(-2.5;0.5^2)}&&\end{array}$\\\\
$w_1=w_2=w_3=1$&$\left\{\begin{array}{ccccccc}x&0.1&0.1231&0.1875&0.5259&0.5922&1\\
n_i&3&2&6&5&1&3\end{array}\right.$\\\\
$\begin{array}{c}w_1=w_2=10;\\w_3=1\end{array}$&$\left\{\begin{array}{ccccccc}x&0.1&0.1267&0.1905&0.5479&0.5962&1\\
n_i&3&2&5&5&1&4\end{array}\right.$\\\\
$\begin{array}{c}w_1=100;\\ w_2=10;~w_3=1\end{array}$&$\left\{\begin{array}{cccccc}x&0.1&0.1202&0.1857&0.5457&1\\
n_i&3&1&6&6&4\end{array}\right.$\\
 \hline
\end{tabular}
\end{small}
\end{center}
\end{table}

Table \ref{tabw2} presents the optimum exact designs for
several priors for $\gamma_1$ and $\gamma_{11}$ and weight
patterns. Now the relation between the variances of the estimators
and the true value of the parameters is much more complicated than
for first order models. These results show the weight approach
used agrees with what is intended in the construction of the
design.

\subsection{Exact optimum design and alternative designs} 

\begin{table}
\caption{Possible designs for the first order FP model. Optimal
design obtained considering $w_1=w_2$, $\gamma_1\sim
N(2.5;1.5^2)$,
$P(\alpha)=(0.15,~0.25,~0.25,~0.15,~0.10,~0.07,~0.03)$,
$\alpha=(-2,~-1,~-1/2,~0,~1/2,~1,~2)$, the design space in the
range $[0.1;1]$, $n=12$ and 200 prior sample points. }
\label{tabfo1a}
\begin{center} {\small
\begin{tabular}{cccccccc}
\hline {\it Design type}&\multicolumn{1}{c}{\it Design}&Efficiency\\
\hline
True Prior&$\left\{\begin{array}{ccccc}x_i&0.1&0.1895 &0.5225&  1\\
n_i&3&4&2&3\end{array}\right.$&100\\\\
 $\begin{array}{c} \mbox{ \small 4 levels}\\\alpha=-1/2\end{array}$&  $\left\{\begin{array}{cccccc}x_i&0.1& 0.1678&0.3377 & 1\\
 n_i&3&3&3&3\end{array}\right.$&93.90\\\\
 $\begin{array}{c} \mbox{ \small 4 levels}\\\alpha=0\end{array}$& $\left\{\begin{array}{cccccc}x_i&0.1& 0.2154&0.4642 & 1\\
 n_i&3&3&3&3\end{array}\right.$&91.50\\\\
 $\begin{array}{c} \mbox{ \small 3 levels}\\\alpha=-1/2\end{array}$&$\left\{\begin{array}{cccc}x_i&0.1& 0.2309&  1\\
 n_i&4&4&4\end{array}\right.$&81.32\\\\
 $\begin{array}{c} \mbox{ \small 4 levels}\\\alpha=-1\end{array}$& $\left\{\begin{array}{cccccc}x_i&0.1& 0.1429&0.2500 & 1\\
 n_i&3&3&3&3\end{array}\right.$&79.69\\\\
 $\begin{array}{c} \mbox{ \small   Point Prior}\\\alpha=-1/2\end{array}$&$\left\{\begin{array}{cccc}x_i&0.1&0.2500 & 1\\
 n_i&3&6&3\end{array}\right.$&78.86\\\\
 $\begin{array}{c} \mbox{ \small 3-lev CCD}\\ \mbox{ \small projection}\\ \alpha=-1/2\end{array}$&     $\left\{\begin{array}{ccccc}x_i&0.1& 0.2309&  1\\
 n_i&3&6&3\end{array}\right.$&72.83\\\\
 $\begin{array}{c} \mbox{ \small 3 levels}\\\alpha=0\end{array}$&$\left\{\begin{array}{cccc}x_i&0.1& 0.3162&  1\\
 n_i&4&4&4\end{array}\right.$&72.47\\\\
$\begin{array}{c} \mbox{ \small 5-lev CCD}\\ \mbox{ \small projection}\\ \alpha=-1/2\end{array}$&     $\left\{\begin{array}{cccccc}x_i&0.1&0.1235&0.2309&0.5768 & 1\\
n_i&1&2&6&2&1\end{array}\right.$&71.58\\\\
 $\begin{array}{c} \mbox{ \small 5-lev CCD}\\ \mbox{ \small projection}\\ \alpha=-1\end{array}$&     $\left\{\begin{array}{cccccc}x_i&0.1&0.1152&0.1818&0.4314 & 1\\
 n_i&1&2&6&2&1\end{array}\right.$&69.91\\
\hline
\end{tabular}}
\end{center}
\end{table}

\begin{table}
\caption{Possible designs for the first order FP model. Optimal
design obtained considering $w_1=w_2$, $\gamma_1\sim
N(2.5;1.5^2)$,
$P(\alpha)=(0.15,~0.25,~0.25,~0.15,~0.10,~0.07,~0.03)$,
$\alpha=(-2,~-1,~-1/2,~0,~1/2,~1,~2)$, the design space in the
range $[0.1;1]$, $n=12$ and 200 prior sample points (continued). }
\label{tabfo1b} \begin{center}{\small
\begin{tabular}{cccccccc}
\hline {\it Design type}&\multicolumn{1}{c}{\it Design}&Efficiency\\
\hline
 $\begin{array}{c} \mbox{ \small Point Prior}\\\alpha=-1\end{array}$&$\left\{\begin{array}{cccc}x_i&0.1&0.2039 & 1\\
 n_i&3&6&3\end{array}\right.$&68.16\\\\
  $\begin{array}{c} \mbox{ \small 5-lev CCD}\\ \mbox{ \small projection}\\ \alpha=0\end{array}$&     $\left\{\begin{array}{cccccc}x_i&0.1&0.1401&0.3162&0.7138 & 1\\
  n_i&1&2&6&2&1\end{array}\right.$&67.13\\\\
$\begin{array}{c} \mbox{ \small Point Prior}\\\alpha=0\end{array}$ &$\left\{\begin{array}{cccc}x_i&0.1 &0.312&  1\\
n_i&5&3&4\end{array}\right.$&66.84\\\\
 $\begin{array}{c} \mbox{ \small 3-lev CCD}\\ \mbox{ \small projection}\\ \alpha=0\end{array}$&     $\left\{\begin{array}{ccccc}x_i&0.1& 0.3162&  1\\
 n_i&3&6&3\end{array}\right.$&62.56\\\\
$\begin{array}{c} \mbox{ \small 3 levels}\\\alpha=-1\end{array}$&$\left\{\begin{array}{cccc}x_i&0.1& 0.1818&  1\\
n_i&4&4&4\end{array}\right.$&56.64\\\\
$\begin{array}{c} \mbox{\small 5-lev CCD}\\\mbox{\small projection}\end{array}$&$\left\{\begin{array}{cccccc}x_i&0.1&0.232 &0.550&0.868 & 1\\
n_i&1&2&6&2&1\end{array}\right.$&53.54\\\\
 $\begin{array}{c} \mbox{ \small 3-lev CCD}\\ \mbox{ \small projection}\\ \alpha=-1\end{array}$&     $\left\{\begin{array}{ccccc}x_i&0.1& 0.1818&  1\\
 n_i&3&6&3\end{array}\right.$&45.04\\\\
 4 levels&$\left\{\begin{array}{ccccc}x_i&0.1& 0.4& 0.7& 1\\
 n_i&3&3&3&3\end{array}\right.$&28.46\\\\
$\begin{array}{c} \mbox{\small 3-lev CCD}\\\mbox{\small projection}\end{array}$&$\left\{\begin{array}{ccccc}x_i&0.1& 0.55&  1\\
n_i&3&6&3\end{array}\right.$&11.23\\\\
 3 levels &$\left\{\begin{array}{cccc}x_i&0.1&0.55&  1\\
 n_i&4&4&4\end{array}\right.$&11.09\\
 \hline
\end{tabular}}
\end{center}
\end{table}

\begin{table}
\caption{Possible designs for the first order FP model. Optimal
design obtained considering $w_1=w_2$, $\gamma_1\sim
N(2.5;1.5^2)$,
$P(\alpha)=(0.03,~0.07,~0.10,~0.15,~0.25,~0.25,~0.15)$,
$\alpha=(-2,~-1,~-1/2,~0,~1/2,~1,~2)$, the design space in the
range $[0.1;1]$, $n=12$ and 200 prior sample points. }
\label{tabfo2a}\begin{center}{\small
\begin{tabular}{cccccccc}
\hline {\it Design type}&\multicolumn{1}{c}{\it Design}&Efficiency\\
\hline True Prior&$\left\{\begin{array}{ccccc}x_i&0.1&0.1912 &0.5381&  1\\
n_i&3&2&4&3\end{array}\right.$&100\\\\
  $\begin{array}{c} \mbox{ \small 4 levels}\\\alpha=1/2\end{array}$&  $\left\{\begin{array}{ccccccc}x_i&0.1& 0.2961&0.5961 & 1\\
  n_i&3&3&3&3\end{array}\right.$&91.95\\\\
   $\begin{array}{c} \mbox{ \small 4 levels}\\\alpha=0\end{array}$&  $\left\{\begin{array}{ccccc}x_i&0.1& 0.2154&0.4642 & 1\\
   n_i&3&3&3&3\end{array}\right.$&87.50\\\\
 $\begin{array}{c} \mbox{ \small   Point Prior}\\\alpha=1/2\end{array}$&$\left\{\begin{array}{cccc}x_i&0.1&0.4006 & 1\\
 n_i&3&6&3\end{array}\right.$&73.17\\\\
 4 levels&$\left\{\begin{array}{ccccc}x_i&0.1& 0.4& 0.7& 1\\
 n_i&3&3&3&3\end{array}\right.$&71.55\\\\
  $\begin{array}{c} \mbox{ \small 5-lev CCD}\\ \mbox{ \small projection}\\ \alpha=1/2\end{array}$& $\left\{\begin{array}{cccccc}x_i&0.1&0.1734 &0.4331&0.8098 & 1\\
  n_i&1&2&6&2&1\end{array}\right.$&71.34\\\\
  $\begin{array}{c} \mbox{ \small 3-lev CCD}\\ \mbox{ \small projection}\\ \alpha=1/2\end{array}$&$\left\{\begin{array}{cccc}x_i&0.1& 0.4331&  1\\
  n_i&3&6&3\end{array}\right.$&70.71\\\\
$\begin{array}{c} \mbox{\small 5-lev CCD}\\\mbox{\small projection}\end{array}$&$\left\{\begin{array}{cccccc}x_i&0.1&0.232 &0.550&0.868 & 1\\
n_i&1&2&6&2&1\end{array}\right.$&70.36\\\\
  $\begin{array}{c} \mbox{ \small 3 levels}\\\alpha=1/2\end{array}$&$\left\{\begin{array}{cccc}x_i&0.1& 0.4331&  1\\
  n_i&4&4&4\end{array}\right.$&67.81\\\\
    $\begin{array}{c} \mbox{ \small 5-lev CCD}\\ \mbox{ \small projection}\\ \alpha=0\end{array}$&     $\left\{\begin{array}{cccccc}x_i&0.1&0.1401&0.3162&0.7138 & 1\\
    n_i&1&2&6&2&1\end{array}\right.$&66.99\\
  \hline
\end{tabular}}
\end{center}
\end{table}

\begin{table}
\caption{Possible designs for the first order FP model. Optimal
design obtained considering $w_1=w_2$, $\gamma_1\sim
N(2.5;1.5^2)$,
$P(\alpha)=(0.03,~0.07,~0.10,~0.15,~0.25,~0.25,~0.15)$,
$\alpha=(-2,~-1,~-1/2,~0,~1/2,~1,~2)$, the design space in the
range $[0.1;1]$, $n=12$ and 200 prior sample points (continued). }
\label{tabfo2b}\begin{center}{\small
\begin{tabular}{cccccccc}
\hline {\it Design type}&\multicolumn{1}{c}{\it Design}&Efficiency\\
\hline
   $\begin{array}{c} \mbox{ \small 3-lev CCD}\\ \mbox{ \small projection}\\ \alpha=0\end{array}$&     $\left\{\begin{array}{ccccc}x_i&0.1& 0.3162&  1\\
   n_i&3&6&3\end{array}\right.$&62.89\\\\
 $\begin{array}{c} \mbox{ \small Point Prior}\\\alpha=1\end{array}$&$\left\{\begin{array}{cccc}x_i&0.1&0.4905 & 1\\
 n_i&3&6&3\end{array}\right.$&59.26\\\\
$\begin{array}{c} \mbox{ \small Point Prior}\\\alpha=0\end{array}$&$\left\{\begin{array}{cccc}x_i&0.1& 0.3073&  1\\
n_i&4&5&3\end{array}\right.$&59.22\\\\
 3 levels &$\left\{\begin{array}{cccc}x_i&0.1& 0.55&  1\\
 n_i&4&4&4\end{array}\right.$&42.53\\\\
$\begin{array}{c} \mbox{\small 3-lev CCD}\\\mbox{\small projection}\end{array}$&$\left\{\begin{array}{cccc}x_i&0.1 &0.55 & 1\\
n_i&3&6&3\end{array}\right.$&43.87\\\\
 $\begin{array}{c} \mbox{ \small 3 levels}\\\alpha=0\end{array}$&$\left\{\begin{array}{cccc}x_i&0.1& 0.3162&  1\\
 n_i&4&4&4\end{array}\right.$&57.20\\\\
  \hline
\end{tabular}}
\end{center}
\end{table}

\begin{table}
\caption{Possible designs for the first order FP model. Optimal
design obtained considering $w_1=w_2$, $\gamma_1\sim
N(2.5;1.5^2)$, $P(\alpha)=(0.05,~ 0.10,~ 0.20,~ 0.30,~ 0.20,~
0.10,~ 0.05 )$, $\alpha=(-2,~-1,~-1/2,~0,~1/2,~1,~2)$, the design
space in the range $[0.1;1]$, $n=12$ and 200 prior sample points.
} \label{tabfo3a}\begin{center}{\small
\begin{tabular}{cccccccc}
\hline {\it Design type}&\multicolumn{1}{c}{\it Design}&Efficiency\\
\hline
True Prior&$\left\{\begin{array}{ccccc}x_i&0.1&0.2000 &0.4703&  1\\
n_i&3&3&3&3\end{array}\right.$&100\\\\
 $\begin{array}{c} \mbox{ \small 4 levels}\\\alpha=0\end{array}$&   $\left\{\begin{array}{ccccc}x_i&0.1& 0.2154&0.4642 & 1\\
 n_i&3&3&3&3\end{array}\right.$&99.25\\\\
$\begin{array}{c} \mbox{ \small 4 levels}\\\alpha=-1/2\end{array}$&  $\left\{\begin{array}{ccccc}x_i&0.1& 0.1678&0.3677 & 1\\
n_i&3&3&3&3\end{array}\right.$&92.93\\\\
$\begin{array}{c} \mbox{ \small 4 levels}\\\alpha=1/2\end{array}$&  $\left\{\begin{array}{ccccc}x_i&0.1& 0.2961&0.5961 & 1\\
n_i&3&3&3&3\end{array}\right.$&81.29\\\\
 $\begin{array}{c} \mbox{ \small Point Prior}\\\alpha=0\end{array}$&$\left\{\begin{array}{cccc}x_i&0.1& 0.3073&  1\\
 n_i&4&5&3\end{array}\right.$&80.45\\\\
 $\begin{array}{c} \mbox{ \small 3-lev CCD}\\ \mbox{ \small projection}\\ \alpha=0\end{array}$&     $\left\{\begin{array}{cccc}x_i&0.1& 0.3162&  1\\
 n_i&3&6&3\end{array}\right.$&80.35\\\\
 4 levels&$\left\{\begin{array}{ccccc}x_i&0.1& 0.4& 0.7& 1\\
 n_i&3&3&3&3\end{array}\right.$&78.45\\\\
 $\begin{array}{c} \mbox{ \small 3 levels}\\\alpha=0\end{array}$&$\left\{\begin{array}{cccc}x_i&0.1& 0.3162&  1\\
 n_i&4&4&4\end{array}\right.$&77.09\\\\
 $\begin{array}{c} \mbox{ \small 5-lev CCD projection}\\ \alpha=0\end{array}$&     $\left\{\begin{array}{cccccc}x_i&0.1&0.1401&0.3162&0.7138 & 1\\
 n_i&1&2&6&2&1\end{array}\right.$&72.31\\\\
$\begin{array}{c} \mbox{ \small 5-lev CCD}\\ \mbox{ \small projection}\\ \alpha=-1/2\end{array}$&     $\left\{\begin{array}{cccccc}x_i&0.1&0.1235&0.2309&0.5768 & 1\\
n_i&1&2&6&2&1\end{array}\right.$&71.44\\
\hline
\end{tabular}}
\end{center}
\end{table}

\begin{table}
\caption{Possible designs for the first order FP model. Optimal
design obtained considering $w_1=w_2$, $\gamma_1\sim
N(2.5;1.5^2)$, $P(\alpha)=(0.05,~ 0.10,~ 0.20,~ 0.30,~ 0.20,~
0.10,~ 0.05 )$, $\alpha=(-2,~-1,~-1/2,~0,~1/2,~1,~2)$, the design
space in the range $[0.1;1]$, $n=12$ and 200 prior sample points
(continued). } \label{tabfo3b}\begin{center}{\small
\begin{tabular}{cccccccc}
\hline {\it Design type}&\multicolumn{1}{c}{\it Design}&Efficiency\\
\hline
  $\begin{array}{c} \mbox{ \small 5-lev CCD}\\ \mbox{ \small projection}\\ \alpha=1/2\end{array}$& $\left\{\begin{array}{cccccc}x_i&0.1&0.1734 &0.4331&0.8098 & 1\\
  n_i&1&2&6&2&1\end{array}\right.$&70.65\\\\
 $\begin{array}{c} \mbox{ \small   Point Prior}\\\alpha=-1/2\end{array}$&$\left\{\begin{array}{cccc}x_i&0.1&0.2500 & 1\\
 n_i&3&6&3\end{array}\right.$&70.24\\\\
$\begin{array}{c} \mbox{\small 5-lev CCD}\\\mbox{\small projection}\end{array}$&$\left\{\begin{array}{cccccc}x_i&0.1&0.232 &0.550&0.868 & 1\\
n_i&1&2&6&2&1\end{array}\right.$&63.79\\\\
 $\begin{array}{c} \mbox{ \small 3-lev CCD}\\ \mbox{ \small projection}\\ \alpha=-1/2\end{array}$&     $\left\{\begin{array}{ccccc}x_i&0.1& 0.2309&  1\\
 n_i&3&6&3\end{array}\right.$&61.06\\\\
 $\begin{array}{c} \mbox{ \small Point Prior}\\\alpha=1/2\end{array}$&$\left\{\begin{array}{cccc}x_i&0.1& 0.4006&  1\\
 n_i&4&5&3\end{array}\right.$&60.64\\\\
$\begin{array}{c} \mbox{ \small 3 levels}\\\alpha=-1/2\end{array}$&$\left\{\begin{array}{cccc}x_i&0.1& 0.2309&  1\\
n_i&4&4&4\end{array}\right.$&58.95\\\\
  $\begin{array}{c} \mbox{ \small 3-lev CCD}\\ \mbox{ \small projection}\\ \alpha=1/2\end{array}$&$\left\{\begin{array}{cccc}x_i&0.1& 0.4331&  1\\
  n_i&3&6&3\end{array}\right.$&50.78\\\\
$\begin{array}{c} \mbox{ \small 3 levels}\\\alpha=1/2\end{array}$&$\left\{\begin{array}{cccc}x_i&0.1& 0.4331&  1\\
n_i&4&4&4\end{array}\right.$&49.25\\\\
$\begin{array}{c} \mbox{\small 3-lev CCD}\\\mbox{\small projection}\end{array}$&$\left\{\begin{array}{cccc}x_i&0.1& 0.55&  1\\
n_i&3&6&3\end{array}\right.$&23.01\\\\
 3 levels &$\left\{\begin{array}{cccc}x_i&0.1& 0.55&  1\\  n_i&4&4&4\end{array}\right.$&22.58\\
 \hline
\end{tabular}}
\end{center}
\end{table}
\begin{table}
\caption{Possible designs for the second order FP model. Optimal
design obtained considering $w_1=w_2=w_3$, $\gamma_1\sim
N(1.0;0.5^2)$, $\gamma_{11}\sim N(-2.5;1.5^2)$,
$P(\alpha)=(0.15,~0.25,~0.25,~0.15,~0.10,~0.07,~0.03)$,
$\alpha=(-2,~-1,~-1/2,~0,~1/2,~1,~2)$, the design space in the
range $[0.1;1]$, $n=20$ and 100 prior sample points. }
\label{tabso1a}\begin{center}{\small
\begin{tabular}{cccccccc}
\hline {\it Design type}&\multicolumn{1}{c}{\it Design}&Efficiency\\
\hline
True Prior&$\left\{\begin{array}{ccccccc}x_i&0.1&0.1400 &0.1742&0.5100 &0.6200& 1\\
n_i&3&1&5&6&1&4\end{array}\right.$&100\\\\
$\begin{array}{c} \mbox{ \small Point Prior}\\\alpha=0\end{array}$ &$\left\{\begin{array}{ccccc}x_i&0.1& 0.1507&0.6073 & 1\\
n_i&2&7&7&4\end{array}\right.$&86.25\\\\
 $\begin{array}{c} \mbox{ \small 5 levels}\\\alpha=0\end{array}$&$\left\{\begin{array}{cccccc}x_i&0.1&01778 &0.3162&0.5623 & 1\\
 n_i&4&4&4&4&4\end{array}\right.$&78.57\\\\
 $\begin{array}{c} \mbox{ \small 5 levels}\\\alpha=-1/2\end{array}$&$\left\{\begin{array}{cccccc}x_i&0.1&0.1455 &0.2309&0.4213 & 1\\
 n_i&4&4&4&4&4\end{array}\right.$&69.57\\\\
 $\begin{array}{c} \mbox{ \small 4 levels}\\\alpha=0\end{array}$& $\left\{\begin{array}{ccccc}x_i&0.1& 0.2154&0.4642 & 1\\
 n_i&5&5&5&5\end{array}\right.$&65.32\\\\
 $\begin{array}{c} \mbox{ \small   Point Prior}\\\alpha=-1/2\end{array}$&$\left\{\begin{array}{ccccc}x_i&0.1&0.1532 &0.4492& 1\\
 n_i&6&4&4&6\end{array}\right.$&60.86\\\\
 $\begin{array}{c} \mbox{ \small 5-lev CCD}\\ \mbox{ \small projection}\\ \alpha=-1\end{array}$& $\left\{\begin{array}{cccccc}x_i&0.1&0.1152&0.1818&0.4314 & 1\\
 n_i&1&4&10&4&1\end{array}\right.$&55.05\\\\
$\begin{array}{c} \mbox{\small 5-lev CCD}\\\mbox{\small projection}\end{array}$&$\left\{\begin{array}{cccccc}x_i&0.1&0.232 &0.550&0.868 & 1\\
n_i&1&4&10&4&1\end{array}\right.$&51.85\\\\
 $\begin{array}{c} \mbox{ \small 4 levels}\\\alpha=-1/2\end{array}$&  $\left\{\begin{array}{ccccc}x_i&0.1& 0.1678&0.3377 & 1\\
 n_i&5&5&5&5\end{array}\right.$&49.71\\\\
 $\begin{array}{c} \mbox{ \small 5-lev CCD}\\ \mbox{ \small projection}\\ \alpha=-1/2\end{array}$& $\left\{\begin{array}{cccccc}x_i&0.1&0.1235&0.2309&0.5768 & 1\\
 n_i&1&4&10&4&1\end{array}\right.$&39.35\\
 \hline
\end{tabular}}
\end{center}
\end{table}
\begin{table}
\caption{Possible designs for the second order FP model. Optimal
design obtained considering $w_1=w_2=w_3$, $\gamma_1\sim
N(1.0;0.5^2)$, $\gamma_{11}\sim N(-2.5;1.5^2)$,
$P(\alpha)=(0.15,~0.25,~0.25,~0.15,~0.10,~0.07,~0.03)$,
$\alpha=(-2,~-1,~-1/2,~0,~1/2,~1,~2)$, the design space in the
range $[0.1;1]$, $n=20$ and 100 prior sample points (continued). }
\label{tabso1b}\begin{center}{\small
\begin{tabular}{cccccccc}
\hline {\it Design type}&\multicolumn{1}{c}{\it Design}&Efficiency\\
\hline
$\begin{array}{c} \mbox{ \small 5 levels}\\\alpha=-1\end{array}$&$\left\{\begin{array}{cccccc}x_i&0.1&0.1290 &0.1818&0.3077 & 1\\
n_i&4&4&4&4&4\end{array}\right.$&36.66\\\\
 $\begin{array}{c} \mbox{ \small 5-lev CCD}\\ \mbox{ \small projection}\\ \alpha=0\end{array}$& $\left\{\begin{array}{cccccc}x_i&0.1&0.1401&0.3162&0.7138 & 1\\
 n_i&1&4&10&4&1\end{array}\right.$&32.96\\\\
$\begin{array}{c} \mbox{ \small Point Prior}\\\alpha=-1\end{array}$&$\left\{\begin{array}{ccccc}x_i&0.1&0.1358 &0.3518& 1\\
n_i&6&3&4&7\end{array}\right.$&30.26\\\\
 5 levels &$\left\{\begin{array}{cccccc}x_i&0.1&0.325 &0.55&0.775 & 1\\
 n_i&4&4&4&4&4\end{array}\right.$&21.77\\\\
 $\begin{array}{c} \mbox{ \small 4 levels}\\\alpha=-1\end{array}$& $\left\{\begin{array}{ccccc}x_i&0.1& 0.1429&0.2500 & 1\\
 n_i&5&5&5&5\end{array}\right.$&17.37\\\\
 4 levels&$\left\{\begin{array}{ccccc}x_i&0.1& 0.4& 0.7& 1\\  n_i&5&5&5&5\end{array}\right.$&7.76\\
 \hline
\end{tabular}}
\end{center}
\end{table}
\begin{table}
\caption{Possible designs for the second order FP model. Optimal
design obtained considering $w_1=w_2=w_3$, $\gamma_1\sim
N(1.0;0.5^2)$, $\gamma_{11}\sim N(-2.5;1.5^2)$,
$P(\alpha)=(0.03,~0.07,~0.10,~0.15,~0.25,~0.25,~0.15)$,
$\alpha=(-2,~-1,~-1/2,~0,~1/2,~1,~2)$, the design space in the
range $[0.1;1]$, $n=20$ and 100 prior sample points. }
\label{tabso2a}\begin{center}{\small
\begin{tabular}{cccccccc}
\hline {\it Design type}&\multicolumn{1}{c}{\it Design}&Efficiency\\
\hline
True Prior&$\left\{\begin{array}{ccccccc}x_i&0.1& 0.1392&0.2199 &0.3941& 0.7619&1\\
n_i&4&2&2&3&4&5\end{array}\right.$&100\\\\
 $\begin{array}{c} \mbox{ \small 5 levels}\\\alpha=1/2\end{array}$&$\left\{\begin{array}{cccccc}x_i&0.1&0.2373 &0.4331&0.6873 & 1\\
 n_i&4&4&4&4&4\end{array}\right.$&71.36\\\\
 $\begin{array}{c} \mbox{ \small 5 levels}\\\alpha=0\end{array}$&$\left\{\begin{array}{cccccc}x_i&0.1&01778 &0.3162&0.5623 & 1\\
 n_i&4&4&4&4&4\end{array}\right.$&70.09\\\\
 $\begin{array}{c} \mbox{ \small 5-lev CCD}\\ \mbox{ \small projection}\\ \alpha=0\end{array}$& $\left\{\begin{array}{cccccc}x_i&0.1&0.1401&0.3162&0.7138 & 1\\
 n_i&1&4&10&4&1\end{array}\right.$&48.95\\\\
 $\begin{array}{c} \mbox{ \small 5-lev CCD}\\ \mbox{ \small projection}\\ \alpha=1/2\end{array}$& $\left\{\begin{array}{cccccc}x_i&0.1&0.1734&0.4331&0.8098 & 1\\
 n_i&1&4&10&4&1\end{array}\right.$&42.53\\\\
 $\begin{array}{c} \mbox{ \small   Point Prior}\\\alpha=1/2\end{array}$&$\left\{\begin{array}{ccccc}x_i&0.1&0.2214&0.6567& 1\\
 n_i&7&3&3&7\end{array}\right.$&36.44\\\\
$\begin{array}{c} \mbox{\small 5-lev CCD}\\\mbox{\small projection}\end{array}$&$\left\{\begin{array}{cccccc}x_i&0.1&0.232 &0.550&0.868 & 1\\
n_i&1&4&10&4&1\end{array}\right.$&36.01\\\\
$\begin{array}{c} \mbox{ \small 4 levels}\\\alpha=1/2\end{array}$&  $\left\{\begin{array}{ccccc}x_i&0.1& 0.2961&0.5961 & 1\\
n_i&5&5&5&5\end{array}\right.$&33.89\\\\
 $\begin{array}{c} \mbox{ \small 4 levels}\\\alpha=0\end{array}$& $\left\{\begin{array}{ccccc}x_i&0.1& 0.2154&0.4642 & 1\\
 n_i&5&5&5&5\end{array}\right.$&33.51\\
 \hline
\end{tabular}}
\end{center}
\end{table}

\begin{table}
\caption{Possible designs for the second order FP model. Optimal
design obtained considering $w_1=w_2=w_3$, $\gamma_1\sim
N(1.0;0.5^2)$, $\gamma_{11}\sim N(-2.5;1.5^2)$,
$P(\alpha)=(0.03,~0.07,~0.10,~0.15,~0.25,~0.25,~0.15)$,
$\alpha=(-2,~-1,~-1/2,~0,~1/2,~1,~2)$, the design space in the
range $[0.1;1]$, $n=20$ and 100 prior sample points (continued). }
\label{tabso2b}\begin{center}{\small
\begin{tabular}{cccccccc}
\hline {\it Design type}&\multicolumn{1}{c}{\it Design}&Efficiency\\
\hline
 5 levels &$\left\{\begin{array}{cccccc}x_i&0.1&0.325 &0.55&0.775 & 1\\
  n_i&4&4&4&4&4\end{array}\right.$&26.85\\\\
$\begin{array}{c} \mbox{ \small Point Prior}\\\alpha=1\end{array}$&$\left\{\begin{array}{ccccc}x_i&0.1&0.2775 &0.7281& 1\\
n_i&7&4&3&6\end{array}\right.$&16.02\\\\
$\begin{array}{c} \mbox{ \small Point Prior}\\\alpha=0\end{array}$ &$\left\{\begin{array}{ccccc}x_i&0.1& 0.1507&0.6073&  1\\
n_i&2&7&7&4\end{array}\right.$&13.72\\\\
 4 levels&$\left\{\begin{array}{ccccc}x_i&0.1& 0.4& 0.7& 1\\
 n_i&5&5&5&5\end{array}\right.$&7.09\\
\hline
\end{tabular}}
\end{center}
\end{table}

\begin{table}
\caption{Possible designs for the second order FP model. Optimal
design obtained considering $w_1=w_2=w_3$, $\gamma_1\sim
N(1.0;0.5^2)$, $\gamma_{11}\sim N(-2.5;1.5^2)$,
$P(\alpha)=(0.05,~0.10,~0.20,~0.30,~0.20,~0.10,~0.05)$,
$\alpha=(-2,~-1,~-1/2,~0,~1/2,~1,~2)$, the design space in the
range $[0.1;1]$, $n=20$ and 100 prior sample points. }
\label{tabso3a}\begin{center}{\small
\begin{tabular}{cccccccc}
\hline {\it Design type}&\multicolumn{1}{c}{\it Design}&Efficiency\\
\hline
True Prior&$\left\{\begin{array}{ccccccc}x_i&0.1& 0.1330&0.1839&0.4368& 0.7671&1\\
n_i&4&2&3&4&3&4\end{array}\right.$&100\\\\
 $\begin{array}{c} \mbox{ \small 5 levels}\\\alpha=0\end{array}$&$\left\{\begin{array}{cccccc}x_i&0.1&01778 &0.3162&0.5623 & 1\\
 n_i&4&4&4&4&4\end{array}\right.$&78.40\\\\
 $\begin{array}{c} \mbox{ \small 5 levels}\\\alpha=1/2\end{array}$&$\left\{\begin{array}{cccccc}x_i&0.1&0.2373 &0.4331&0.6873 & 1\\
 n_i&4&4&4&4&4\end{array}\right.$&66.13\\\\
 $\begin{array}{c} \mbox{ \small 5-lev CCD}\\ \mbox{ \small  projection}\\ \alpha=0\end{array}$& $\left\{\begin{array}{cccccc}x_i&0.1&0.1401&0.3162&0.7138 & 1\\
 n_i&1&4&10&4&1\end{array}\right.$&48.95\\\\
 $\begin{array}{c} \mbox{ \small 5-lev CCD}\\ \mbox{ \small projection}\\ \alpha=1/2\end{array}$& $\left\{\begin{array}{cccccc}x_i&0.1&0.1734&0.4331&0.8098 & 1\\
 n_i&1&4&10&4&1\end{array}\right.$&47.87\\\\
 $\begin{array}{c} \mbox{ \small 4 levels}\\\alpha=0\end{array}$& $\left\{\begin{array}{ccccc}x_i&0.1& 0.2154&0.4642 & 1\\
 n_i&5&5&5&5\end{array}\right.$&42.88\\\\
 $\begin{array}{c} \mbox{ \small 5-lev CCD}\\ \mbox{ \small projection}\\ \alpha=-1/2\end{array}$& $\left\{\begin{array}{cccccc}x_i&0.1&0.1235&0.2309&0.5768 & 1\\
 n_i&1&4&10&4&1\end{array}\right.$&41.73\\\\
 $\begin{array}{c} \mbox{5-lev CCD}\\ \mbox{ \small projection}\end{array}$&$\left\{\begin{array}{cccccc}x_i&0.1&0.232 &0.550&0.868 & 1\\
 n_i&1&4&10&4&1\end{array}\right.$&38.30\\\\
 $\begin{array}{c} \mbox{ \small 5 levels}\\\alpha=-1/2\end{array}$&$\left\{\begin{array}{cccccc}x_i&0.1&0.1455 &0.2309&0.4213 & 1\\
 n_i&4&4&4&4&4\end{array}\right.$&34.74\\\\
 $\begin{array}{c} \mbox{ \small   Point Prior}\\\alpha=1/2\end{array}$&$\left\{\begin{array}{ccccc}x_i&0.1&0.2214&0.6567& 1\\
 n_i&7&3&3&7\end{array}\right.$&34.31\\
 \hline
\end{tabular}}
\end{center}
\end{table}
\begin{table}
\caption{Possible designs for the second order FP model. Optimal
design obtained considering $w_1=w_2=w_3$, $\gamma_1\sim
N(1.0;0.5^2)$, $\gamma_{11}\sim N(-2.5;1.5^2)$,
$P(\alpha)=(0.05,~0.10,~0.20,~0.30,~0.20,~0.10,~0.05)$,
$\alpha=(-2,~-1,~-1/2,~0,~1/2,~1,~2)$, the design space in the
range $[0.1;1]$, $n=20$ and 100 prior sample points (continued). }
\label{tabso3b}\begin{center}{\small
\begin{tabular}{cccccccc}
\hline {\it Design type}&\multicolumn{1}{c}{\it Design}&Efficiency\\
\hline
$\begin{array}{c} \mbox{ \small 4 levels}\\\alpha=1/2\end{array}$&  $\left\{\begin{array}{ccccc}x_i&0.1& 0.2961&0.5961 & 1\\
n_i&5&5&5&5\end{array}\right.$&29.89\\\\
  5 levels &$\left\{\begin{array}{cccccc}x_i&0.1&0.325 &0.55&0.775 & 1\\
  n_i&4&4&4&4&4\end{array}\right.$&22.50\\\\
$\begin{array}{c} \mbox{ \small Point Prior}\\\alpha=0\end{array}$ &$\left\{\begin{array}{ccccc}x_i&0.1& 0.1507&0.6073&  1\\
n_i&2&7&7&4\end{array}\right.$&20.13\\\\
$\begin{array}{c} \mbox{ \small Point Prior}\\\alpha=-1/2\end{array}$&$\left\{\begin{array}{ccccc}x_i&0.1&0.1532 &0.4492& 1\\
n_i&7&4&3&6\end{array}\right.$&12.54\\\\
 $\begin{array}{c} \mbox{ \small 4 levels}\\\alpha=-1/2\end{array}$&  $\left\{\begin{array}{ccccc}x_i&0.1& 0.1678&0.3377 & 1\\
 n_i&5&5&5&5\end{array}\right.$&9.08\\\\
 4 levels&$\left\{\begin{array}{ccccc}x_i&0.1& 0.4& 0.7& 1\\  n_i&5&5&5&5\end{array}\right.$&5.90\\
 \hline
\end{tabular}}
\end{center}
\end{table}

 In this section we present exact optimum
Bayesian designs for some particular cases and compare them to
several popular designs such as equally spaced and equally
replicated levels, one-factor projections of central composite
designs (CCD) and all these in the transformed metric for the best
guess of $\alpha$. We also consider locally optimum designs
setting point prior information for the best guess of each
parameter. Tables \ref{tabfo1a}-\ref{tabfo3b} show designs for
first order ($n=12$) and \ref{tabso1a}-\ref{tabso3b} show
designs for second order FP models ($n=20$). The designs are
ordered in terms of their efficiencies which were calculated with
respect to the Bayesian optimum exact designs. Note that, although
we can not guarantee such designs are the global optima they
performed better that any other design we could think of. As
expected, such designs tend to use more levels than the number of
parameters in the model reflecting their robustness over prior
point or locally optimum designs.

For first order designs, in general, 4 levels equally spaced in
the transformed metric (using the most probable values for
$\alpha$) resulted in reasonable designs, these being better than
prior point designs. In general, one factor CCD projections
performed poorly as well as most 3 level designs.

For second order designs, in general, even the best popular
designs were quite far from the optimum designs. Among these were
5 level designs, equally spaced and CCD projections, in some
transformed metric. Most 4 level and locally optimum designs
performed quite poorly.

\subsection{Two-factor designs}

A brief study was conducted considering two factors. The second order fractional polynomial model is
\[\eta({\bf x},\bftheta)=\beta_0+\beta_1x_1^{(\alpha_1)}+\beta_{11}\left\{x_1^{(\alpha_1)}\right\}^2+ \beta_2x_2^{(\alpha_2)}+\beta_{22}\left\{x_2^{(\alpha_2)}\right\}^2+\beta_{12}x_1^{(\alpha_1)}x_2^{(\alpha_2)},
\]
for $x_1,~x_2>0$. The information matrix $\mathbf{M}(\bftheta, \mathbf{x})$ is $8\times8$ and even $D$-optimum designs depend on the values of the $\beta$ parameters. 

For incorporating prior information on $\beta$s, a similar reparametrization is used as for a single factor in order to get interpretable quantities which are relatively easy for experimenters to express prior information on. For $x_j \in [x_{j0},~1], ~j=1,2$, we use the following expected quantities. The slope for a single factor is measured by
	\begin{eqnarray*} 
		\gamma_k 
		&=& \beta_k(1^{(\alpha_k)}-x_{k0}^{(\alpha_k)})+\beta_{kk}(1^{(\alpha_k)}-x_{k0}^{(\alpha_k)})(1^{(\alpha_k)}+x_{k0}^{(\alpha_k)}) +\\
		& &\frac{1}{2}\beta_{12}(1^{(\alpha_k)}-x_{k0}^{(\alpha_k)})(1^{(\alpha_k^\prime)}+x_{k^\prime0}^{(\alpha_k^\prime)}).
	\end{eqnarray*}
For the curvature of a single factor, we use
\begin{eqnarray*} 
	\gamma_{kk}&=&\frac{1}{4}\beta_{kk}(1^{(\alpha_k)}-x_{k0}^{(\alpha_k)})^2.\end{eqnarray*}
Finally, for the interaction of two factors, we use the difference in slopes of one factor across the range of the other factor, namely
\begin{eqnarray*} 
\gamma_{12}&=&\frac{1}{2}\beta_{12}(1^{(\alpha_1)}-x_{10}^{(\alpha_1)})(1^{(\alpha_{2})}-x_{20}^{(\alpha_2)}).
\end{eqnarray*}	

\begin{figure}\label{fig:fig1}
\includegraphics[width=\textwidth, height=12cm]{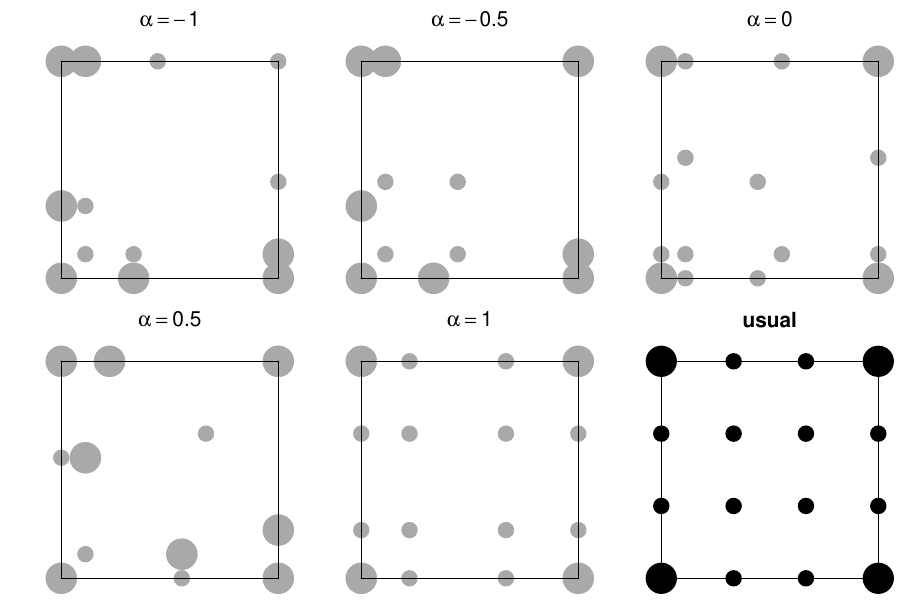}
\caption{Locally $D$-optimum designs for $n=20,~\alpha_1=\alpha_2;~\gamma_1=\gamma_2=\gamma_{12}=1.0;~\gamma_{11}=\gamma_{22}=-2.5$}
\end{figure}

\begin{table}
\begin{center}
\caption{Efficiencies of designs with respect to local $D$-optimal design}
\begin{tabular}{crrrrr}
		\hline
	True	&\multicolumn{5}{c}{Designs ($\alpha_1=\alpha_2$)}\\\cline{2-6}
		($\alpha_1,~\alpha_2$)& \multicolumn{1}{c}{$-1.0$} &\multicolumn{1}{c} {$-0.5$} &\multicolumn{1}{c}{$~~0.0$} & \multicolumn{1}{c}{$+0.5$} &\multicolumn{1}{c} {$+1.0$}\\\hline		
		($-1.0,~-1.0$)  &{100.0} & 98.5 & 90.9 & 80.2 & {50.9} \\ 
		($-0.5,~-0.5$) & 99.4 & {100.0} & 97.3 & 87.1 & {64.9} \\ 
		($~~0.0,~~~~~0.0$)   & 92.8 & 96.7 & {100.0} & 94.4 & 81.2 \\  
		($+0.5,~+0.5$)  & 81.9 & 89.6 & 97.4 &{100.0} & 97.0 \\ 
		($+1.0,~+1.0$)  & {62.9} & 72.4 & 82.4 & 93.6 & {100.0} \\ \hline
		($-1.0,~-0.5$) & 99.8 & 99.2 & 94.1 & 83.3 & {57.5} \\ 
		($-1.0,~~~~~0.0$) & 96.7 & 97.2 & 95.4 & 86.3 & {64.2} \\ 
		($-1.0,~+0.5$) & 91.3 & 92.8 & 94.7 & 88.7 & 70.3 \\ 
		($-1.0,~+1.0$) & 80.5 & 82.9 & 87.5 & 84.9 & 71.4 \\ 
		($-0.5,~~~~~0.0$) & 96.4 & 98.4 & 99.0 & 90.8 & 72.8 \\ 
		($-0.5,~+0.5$) & 90.5 & 93.9 & 97.8 & 92.9 & 79.5 \\ 
		($-0.5,~+1.0$) & 79.7 & 84.1 & 90.3 & 89.2 & 80.7 \\ 
		($~~0.0,~+0.5$) & 87.2 & 92.9 & 98.8 & 96.9 & 88.7 \\ 
		($~~0.0,~+1.0$) & 76.6 & 83.3 & 91.1 & 93.3 & 90.1 \\ 
		($+0.5,~+1.0$) & 72.0 & 80.7 & 89.9 & 96.9 & 98.8 \\ 
		\hline
		mean loss&12.8&9.2 & 6.2 & 9.4&{22.1}\\
		max loss&{37.1} &{27.6}&17.6 &19.8 &{49.1}\\
	\hline
	\end{tabular}
\end{center}
\label{tab:localD}
\end{table}

First we used the coordinate exchange algorithm to obtain the locally $D$-optimum design shown in Figure 1. The local $D$-efficiencies of various designs are shown in Table 15. It can be seen that the designs are robust to small changes in the values of the $\alpha$ parameters, but not to larger changes. Thus the locally optimal design is a good choice if the experimenters are fairly sure about the scale on which a quadratic approximation is appropriate.

\begin{table}
\caption{Priors for Bayesian optimal designs}
\begin{center}
	\begin{tabular}{ccrrrrr}
	\hline
	&&\multicolumn{5}{c}{$\alpha$}\\\cline{3-7}
	Type	&	&-1.0&-0.5&0.0&0.5&1\\\hline
	U$_i$ & & .20&.20&.20&.20&.20\\
	S$_i$ & & .10&.20&.40&.20&.10\\	
	R$_i$ & & .45&.30&.15&.07&.03\\
	L$_i$ & & .03&.07&.15&.30&.45\\		
	\hline\end{tabular}\end{center}
\label{tab:Priors}
\end{table}
    
The pseudo-Bayesian $D$-optimum designs for two-factors were found using the discrete priors for $\alpha$s shown in Table \ref{tab:Priors} and using quadrature to approximate the criterion. The optimal designs were 
similar whether normal priors or point priors were used for the $\gamma$ parameters.

\begin{figure}	\vspace{-.7cm}
\includegraphics[width=\textwidth, height=12cm]{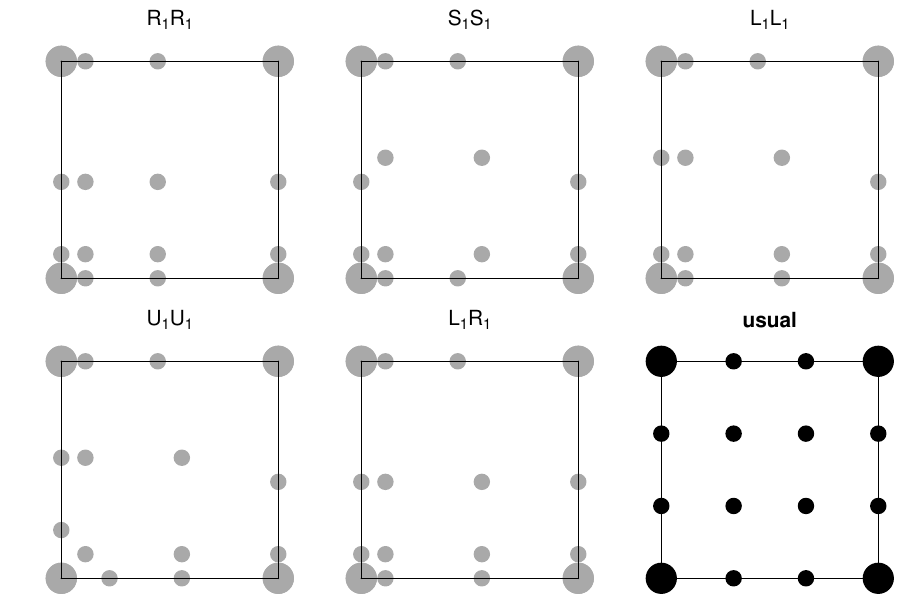}
\caption{Pseudo-Bayesian $D$-optimum designs for $n=20$ $\gamma_1,~\gamma_2,~\gamma_{12}\sim N(1.0,~0.2)$; $\gamma_{11},~\gamma_{22}\sim N(-2.5,~0.5)$}
\label{fig:BayesD}
\end{figure}

\begin{table}
    \centering
        \caption{Bayesian $D$-efficiences of designs}
\begin{tabular}{rrrrrrrrr}
		\hline
		&\multicolumn{8}{c}{Designs}\\\cline{2-9}
		&U$_1$U$_2$&U$_1$S$_2$&U$_1$R$_2$&U$_1$L$_2$&S$_1$S$_2$&R$_1$R$_2$&L$_1$L$_2$&L$_1$R$_2$\\
		Prior& \multicolumn{1}{c}{(1)} &\multicolumn{1}{c} {(2)} &\multicolumn{1}{c}{ (3)} & \multicolumn{1}{c}{(4)} &\multicolumn{1}{c} {(5)} &\multicolumn{1}{c} {(6)} &\multicolumn{1}{c} {(7)} &\multicolumn{1}{c} {(8)} \\ 
		\hline
		U$_1$U$_2$ &{ 100.00}& 99.7 &   98.6 & 97.8 & 99.5 &97.1 & 95.8 & 96.8 \\ 
		U$_1$S$_2$ & 99.8 & { 100.00} &   98.7 & 97.2 & 99.7 &97.2 & 95.6 & 97.1 \\ 
		U$_1$R$_2$ & 98.4 & 98.7 &   { 100.00} & 93.7 & 98.3 & 98.5 & 90.8 & 99.1 \\ 
		U$_1$L$_2$ & 99.2 & 98.0 &    94.6 & { 100.00} & 98.0 &93.1 & 98.7 & 91.9 \\ 
		S$_1$S$_2$ & 99.7 & 99.9 &    98.6 & 96.9 & { 100.00} &97.4 & 95.1 & 96.4 \\ 
		R$_1$R$_2$ & 97.0 & 97.1 &    98.5 & 92.5 & 97.1 &{ 100.00} & 87.1 & 96.2 \\ 
		L$_1$L$_2$ & 98.3 & 97.1 &    93.6 & 99.3 & 96.4 &89.1 &{ 100.00} & 92.9 \\ 
		L$_1$R$_2$ & 97.4 & 97.8 &    99.1 & 92.5 & 96.7 &94.5 & 92.6 & { 100.00}\\ \hline
mean loss & 1.3 & 1.5 &   2.3 & 3.8 &1.8 & 4.2 & 5.6 & 3.7 \\ 
max loss& 3.0 & 2.9 &   6.4 & 7.5 & 3.6 & 10.9 & 12.9 & 8.1 \\ 
		\hline
	\end{tabular}
    \label{tab:BayesEff}
    \end{table}

\begin{table}
    \centering
    \caption{Bayesian efficiencies of locally optimal designs}
\begin{tabular}{crrrrr}\hline
& \multicolumn{5}{c}{Locally ($\alpha_1=\alpha_2=\alpha$)}\\\cline{2-6}
p-Bayesian& \multicolumn{1}{c}{$\alpha=-1$} &\multicolumn{1}{c} {$\alpha=-.5$} &\multicolumn{1}{c}{ $\alpha=0$} & \multicolumn{1}{c}{$\alpha=.5$} &\multicolumn{1}{c} {$\alpha=1$} \\ 
\hline
R$_1$R$_2$&  99.0 & 99.7 & 97.1& 88.0 & {64.7} \\ 
S$_1$S$_2$&  92.6 & 96.6 & 100.0 & 95.6 & 81.5 \\ 
L$_1$L$_2$&  79.9 & 87.6 & 95.7 & 99.6 & 96.4 \\ 
U$_1$U$_2$&  91.8 & 95.9 & 99.3 & 95.8 & 81.2 \\ 
U$_1$S$_2$&  92.2 & 96.2 & 99.6 & 95.6 & 81.3 \\ 
U$_1$R$_2$&  95.4 & 97.6 & 98.3 & 91.7 & 72.4 \\ 
U$_1$L$_2$&  85.8 & 91.6 & 97.6 & 97.6 & 88.5 \\ 
L$_1$R$_2$&  89.3 & 92.8 & 96.7 & 93.1 & 78.9 \\
 \hline
mean loss&   9.3 & 5.3 & 2.0 & 5.4 & {19.4} \\ 
max loss&{20.1} & 12.4 & 4.3 & 12.0 & {35.3} \\ 
	\hline	
\end{tabular}
\label{tab:EffofLocal}
\end{table}

The resulting optimal design is shown in Figure \ref{fig:BayesD} and the efficiencies of designs are shown in Table \ref{tab:BayesEff}. The robustness of the Bayesian designs is immediately apparent, all efficiencies being above $87\%$. The Bayesian efficiencies of locally optimal designs are shown in Table \ref{tab:EffofLocal}, confirming that the locally optimal designs are less robust.

\section{Practical advice for choice of a design}\label{advice}
The results from the last section show that, for one factor, four-level
designs are very good designs in case a power transformation
parameter is to be estimated, the replication pattern being
dependent on the relative interest in the regression coefficients. More
precise designs for the power parameter should use more replicates at
the extreme levels. The space between the levels are quite
dependent on the $\alpha$ prior values and the proposed methodology to
use prior information for the choice of robust designs is
very useful. However in the absence of such methodology equally
spaced and replicated four-level designs in the transformed metric
using the best guess for $\alpha$ would be good alternatives.

For second order FP models the chance of not finding the global
optimum design is greater but we believe in practice the obtained
design is not too inferior to the optimum. We believe in this case
the proposed methodology is very helpful since standard designs
are quite inferior compared to the optimum we found. It should be
noted however that the equally spaced in the log scale
($\alpha=0$) and equally replicated five-level design is a quite
robust design for any of the prior distributions considered,
although in the best situation it was only about 86\% efficient
compared to the optimum.

We believe such methodology can be straightforwardly extended for finding
multi-factor designs, as illustrated by the two-factor example given. However the design search over
multidimensional grids is the main difficulty because of the long
computing time required. Perhaps one alternative would be doing
something similar to the case of standard polynomials where the
candidate design points are based on complete factorials. Perhaps
for FP models, given the results for one factor, we could
construct a set of candidate points based on factorial
combinations using transformed metrics for the best guesses of the
$\alpha$'s. The resulting design could then be improved further by
searching around the design points in a similar way to the adjustment
algorithm of \cite{donev_atkinson_1988}.

\section*{Acknowledgements}

Much of this work was carried out during a number of study visits.
We are grateful for funding from FAPESP grant numbers 01/03151-2,
01/13115-3 and 03/05598-0 and EPSRC grant number EP/T021624/1.

\bibliography{Bibliography}

\begin{thebibliography}{}

\bibitem[Atkinson et~al., 2007]{Atkinson2007}
Atkinson, A.~C., Donev, A.~N., and Tobias, R.~D. (2007).
\newblock {\em Optimum Experimental Designs, with SAS}.
\newblock Oxford: Oxford University Press.

\bibitem[Box and Draper, 1987]{box_draper_1987}
Box, G. E.~P. and Draper, N.~R. (1987).
\newblock {\em Empirical Model-Building and Response Surfaces}.
\newblock Wiley, New York.

\bibitem[Box and Tidwell, 1962]{box_tidwell_1962}
Box, G. E.~P. and Tidwell, P.~W. (1962).
\newblock Transformation of the independent variables.
\newblock {\em Technometrics}, 4:531--550.

\bibitem[Chaloner and Verdinelli, 1995]{chaloner_verdinelli_1995}
Chaloner, K. and Verdinelli, I. (1995).
\newblock Bayesian experimental design: a review.
\newblock {\em Statistical Science}, 10:273--304.

\bibitem[Donev and Atkinson, 1988]{donev_atkinson_1988}
Donev, A.~N. and Atkinson, A.~C. (1988).
\newblock An adjustment algorithm for the construction of exact d-optimum
  experimental designs.
\newblock {\em Technometrics}, 30:429--433.

\bibitem[Gilmour and Trinca, 2005]{gilmour_trinca_2005a}
Gilmour, S.~G. and Trinca, L.~A. (2005).
\newblock Fractional polynomial response surface models.
\newblock {\em Journal of Agricultural, Biological, and Environmental
  Statistics}, 10:50--60.

\bibitem[Gilmour and Trinca, 2012]{gilmour_trinca_2012b}
Gilmour, S.~G. and Trinca, L.~A. (2012).
\newblock Bayesian ${L}$-optimal exact design of experiments for biological
  kinetic models.
\newblock {\em Journal of the Royal Statistical Society. Series C: Applied
  Statistics}, 61:237--251.

\bibitem[Pukelsheim, 1993]{pukelsheim_1993}
Pukelsheim, F. (1993).
\newblock {\em Optimal Design of Experiments}.
\newblock Wiley, New York.

\end{thebibliography}

\end{document}